\documentstyle[12pt]{article}
\begin{document}
\begin{center}
{\bf Effect of randomness and anisotropy on Turing patterns in
reaction-diffusion systems}
\vskip 1cm
Indrani Bose and Indranath Chaudhuri
\\Department of Physics, \\Bose Institute,
\\93/1, Acharya Prafulla Chandra Road,
\\Calcutta-700 009, India.
\end{center}

\begin {abstract}
We study the effect of randomness and anisotropy on Turing patterns in 
reaction-diffusion systems. For this purpose, the Gierer-Meinhardt model of
pattern formation is considered. The cases we study are: (i)randomness in the underlying lattice
structure, (ii)the case in which there is a probablity p that at a lattice site
both reaction and diffusion occur, otherwise there is only diffusion and lastly,
the effect of (iii)anisotropic and(iv)random diffusion coefficients on the formation of 
Turing patterns. The general conclusion is that the Turing mechanism of pattern
formation is fairly robust in the presence of randomness and anisotropy.
\end{abstract}


\section*{I. Introduction}

In 1952 Turing \cite{Turing} pointed out that diffusion need not
always act to smooth out concentration differences in a chemical
system. Two interacting chemicals can generate a stable,  
 inhomogeneous pattern if one of the substances(the inhibitor) 
diffuses much faster than the other(the activator). The
activator is autocatalytic, i.e, a small increase in its concentration
`a' over its homogeneous steady-state concentration leads to a further
increse of `a'. The activator besides promoting its own production
also promotes the production of the inhibitor. The inhibitor, as
the name implies, is antagonistic to the activator and inhibits
its production. Suppose, originally the system is in a
homogeneous steady state. A local increase in the activator concentration
leads to a further increase in the concentration of the activator
due to autocatalysis. The concentration of the inhibitor is also increased
locally. The inhibitor, having a diffusion coefficient much
larger than that of the activator, diffuses faster to the
surrounding region and prevents the activator from coming there. This
process of autocatalysis and long-rang inhibition finally lead
to a stationary state consisting of islands of high activator
concentration within a region of high inhibitor
concentration. The islands constitute what is known as the 
Turing pattern. Turing's original idea was that the stable
patterns could be linked to the patterns seen in biological systems.
Experimental evidence of Turing patterns, however, came much
later and that too not in biological systems but in chemical systems
\cite{Castets,Ouyang,Lee}. This has sparked renewed interest in
mathematical models of pattern formation as well as the relationship
of chemical patterns to the remarkably similar patterns observed in
diverse physical and biological systems \cite{Cross}. Turing structures
have also been seen in electrical gas discharge systems \cite{Astrov}.
Recently, it has been suggested that the formation of stripe patterns
on the marine angelfish Pomacanthus can be explained on the
basis of Turing mechanism involving reaction-diffusion(RD) \cite{Kondo}.
A RD neural network model has been proposed based on
nonsynaptic diffusion neurotransmission \cite{Liang}. The model
has the features of short-range activation and long-rang
inhibition, necessary ingredients for the formation of Turing patterns.
The network exhibits similar self-organization behaviour.

        The diverse examples mentioned above show that the
Turing patterns are not unique to a particular system. Also
Turing mechanism embodies a general principle of self-organization.
A well-known model of a RD system in which Turing patterns can
form is the Gierer-Meinhardt(GM) model \cite{Gierer,Koch}.
In Section II, we describe the GM model and study the effect of 
randomness in the structure of the RD system on the pattern 
formation process(Case I). For this purpose the differential equations
of the model are discretized on a square lattice. We further
study the situation in which there is a probability p that at a
lattice site both reaction and diffusion occur, otherwise there
is only diffusion(Case II). In Section III, we study the effect of 
anisotropic and random diffusion coefficients(Case III and Case IV)
on the formation of Turing patterns. All the studies are based on computer
simulation on a squre lattice. Section IV contains a general discussion
of the models studied.

\section*{II. Randomness in structure and dynamics}

The differential equations describing RD in the GM model are

\begin{eqnarray*}
\frac{\partial a}{\partial t}\,&=&\,D_a\,\Delta a\,+\,\rho_a\,\frac{a^2}{h}\,
-\,\mu_a\,a \quad\quad\quad\quad          (1a)      \\
\frac{\partial h}{\partial t}\,&=&\,D_h\,\Delta h\,+\,\rho_h\,{a^2}\,
-\,\mu_h\,h \quad\quad\quad\quad          (1b)     
\end{eqnarray*}
         where $\Delta $ is the Laplacian given by $\Delta \,=\,\frac{\partial^2}
         {\partial x^2}\,+\,\frac{\partial^2}{\partial y^2}\,$, `a' and `h' denote the concentrations 
of the activator and the inhibitor, $D_a,\,\,D_h$, are the respective diffusion
coefficients, $\mu_a,\,\mu_h$ are the removal rates and $\rho_a,\,\rho_h$
are the cross-reaction coefficients. The conditions for the formation of 
stable Turing patterns are $D_h\,>>D_a$ and $\mu_h\,>\mu_a$ \cite{Koch}.
We also assume that $\rho_a\,=\mu_a$ and $\rho_h\,=\mu_h$. In this case,
the steady state solution of equations (1a) and (1b) is given by (a,h)=(1,1),
i.e., the steady state is homogeneous. The homogeneous steady state is stable  
if local fluctuations created in the system decay with time. If the fluctuations
grow with time, the original homogeneous state is unstable. A phase diagram 
($\mu\,=\,{\bf \frac{\mu_h}{\mu_a}}$ versus $D\,=\,{\bf \frac{D_a}{D_h}}$), based on the
linear stability analysis, of the one-dimensional(1d) version of the GM model
is given in Ref.\cite{Koch}. The phase diagram contains a region in which Turing
patterns can form. In this parameter regime, instability in the original homogeneous
state leads finally to a steady state in which Turing patterns of high activator
concentration are formed. Appendix B of the same Ref.\cite{Koch} gives the parameter
values for a two-dimensional(2d) RD system for which Turing patterns can form 
in the steady state. We use these parameter values, $D_a\,=\,0.005$, $\rho_a\,=\,\mu_a\,=\,0.01$
and $D_h\,=\,0.2$, $\rho_h\,=\,\mu_h\,=\,0.02$, for our studies.

                         As in \cite{Koch} a very simple discretization scheme
is used. The Laplacian $\Delta$ applied to the function a(x,t) is taken as                          
\begin{eqnarray*}
\Delta\,a(x_{ij},t)\,=\,\frac{a(x_{i+1j},t)+a(x_{ij+1},t)+a(x_{i-1j},t)
+a(x_{ij-1},t)-4a(x_{ij},t)}{\delta\,x^2}       \quad\quad\quad\quad\quad (2)
\end{eqnarray*}
where $x_{ij}$ denotes a lattice site,  $x_{ij}\,=\,(i\delta x,\,j\delta x)$.
Time is also discretized, $t_k\,=\,k\delta t$, and the time derivative approximated as
\begin{eqnarray*}    
\frac{\partial a(x,\,t_k)}{\partial t}\,=\,\frac{a(x,\,t_{k+1})\,-\,a(x,\,t_k)}{\delta t}  \quad\quad\quad  (3)
\end{eqnarray*}
In all our simulations, we choose $\delta x\,=\delta t\,=1$. The lattice chosen is of
 size $30\,\times\,30$. Also, periodic boundary conditions are assumed.

              We first study the RD process on inhomogeneous substrata(CaseI). In our
case these are the (2d) percolation clusters, with site occuption probability p,
on which the activator and inhibitor react and dffuse. The percolation clusters
are generated in the usual manner with the help of a random number generator. 
If the random number is less than or equal to p, the site of the lattice is
occupied, otherwise it is kept empty. All the sites of the square lattice are
examined successively and the occupation status of a site determined with the
help of the random number generator. The nearest-neighbour(n.n.) occupied sites
constitute a percolation cluster.

The differential equations (1a) and (1b) are discretized according to the 
schemes specified in equations (2) and (3). The Laplacian in (2) is now written as
\begin{eqnarray*}
\Delta\,a(x_{ij},\,t)\,&=&\,iocc(i+1,\, j)\,\times\,(\,a(x_{i+1j},\, t)-a(x_{ij},\,t)\,)\\
&+&iocc(i,\,j+1)\,\times\,(\,a(x_{ij+1},\,t)-a(x_{ij},\,t)\,)\\
&+&iocc(i-1,\,j)\,\times\,(\,a(x_{i-1j},\,t)-a(x_{ij},\,t)\,)\\
&+&iocc(i,\,j-1)\,\times\,(a(x_{ij-1},\,t)-a(x_{ij},\,t)\,)  \quad\quad\quad\quad      (4)
\end{eqnarray*}
The array iocc keeps track of the occupation status of the sites of the square
lattice. If the site $x_{ij}$ is occupied then iocc(i,j)=1, otherwise it is 
equal to zero. Equation(4) expresses the fact that diffusion to a site from a 
neighbouring site takes place only if the neighbouring site belongs to the RD
network, i.e., to a percolation cluster. One can easily check that the discretized
differential equations (with $\rho_a\,=\,\mu_a$ and $\rho_h\,=\,\mu_h$) have a
steady state solution given by a=1 and h=1 for all the cluster sites.
Random fluctuations of magnitude less than 0.1 are created in the steady state with
the help of the random number generator. This fixes the values of a and h at all
the cluster sites at time t=0. The values of a and h at time t+1 are determined
at a site $x_{ij}$ belonging to a cluster with the help of the discrete equations
for a and h. This process is repeated till the steady state is reached, i.e., the
values of a and h at all the cluster sites do not change within a specified accuracy.

We define an `activated' zone as an island of n.n. sites in the steady state, 
at each of which the activator concentration has a value greater than 1 which is the
homogeneous steady state value. Figs. 1(a)-1(d) show the concentration profiles
in the activated zones for site occupation probabilities p=0.9, 0.7, 0.59, and 0.4 respectively.
The value $ p\,=\,p_c\,=0.59$ is the site percolation threshold for a square
lattice. For $p\,>\,p_c$, a connected network of sites spans the lattice, 
for $p\,<\,p_c$, there is no spanning cluster. For $p\,>\,p_c$, the percolation
clusters include both the spanning cluster as well as isolated clusters in other
parts of the lattice. For $p\,<\,p_c$, the percolation clusters consist of only
the isolated clusters.

The Figures show that the number of activated zones increases as p decreases.
This trend continues below the percolation threshold. Also, the average height
of the concentration peaks decreses. These results can be understood in the 
following manner. With lesser connections in the RD network, as p decreses,
the inhibitor cannot diffuse to long distances and so cannot prevent activator
growth in the local regions. Thus in the steady state there is a larger number of 
activated zones. The average height of activator-concentration profiles decreases
because of the inability of the inhibitor to totally diffuse away from the 
activator zone. The greater concentration of inhibitor in this zone limits the 
growth of the activator concentration more than in the case of a regular RD network.

We next consider CaseII. In this case, the RD network is a fully connected
square lattice. Let p be the probablity that both reaction and diffusion occur 
at a site. The other possibility, with probablity (1-p), is ordinary diffusion.
When p=1, i.e., there is no randomness in the dynamics, Turing patterns form in
the steady state. When p=0, i.e., there is no reaction, simple diffusion takes
over. The steady state in this case is homogeneous with all concentration
gradients removed. Figs. 2(a)-2(d) show the steady state patterns for 
p=0.01, 0.05, 0.3 and 0.7 . For p=0.005, the steady state is homogeneous.
Thus for p as small as 0.01, i.e., when RD occurs at only 1\% of the lattice
sites a Turing pattern is formed. In this case, there is only one activated
zone which covers a large area. As p increses, the number of activated zones increses.
For small values of p, we have checked that an activated zone need not be centred 
around a cluster of n.n. sites at which RD occurs, a zone may form in the intermediate
region of two such clusters. In the light of this fact, it is interesting to note
that even with very few RD sites, the Turing mechanism is operative.

\section*{III. Anisotropy and randomness in diffusion coefficients}

Anisotropy in the RD medium is reflected in the anisotropy of the diffusion 
coefficients. Mertens et al \cite{Mertens} have studied the effect of anisotropic 
diffusion coefficients on pattern formation in catalytic surface reactions.
Their conclusion is that anisotropy may give rise to new types of patterns.
In our Case III, we assume that for diffusion in the vertical direction,
the diffusion coefficients are $D_a\,=0.005$ and $D_h\,=0.2$ . These are the values 
for which Turing patterns can form in an isotropic medium. For diffusion in the
horizontal direction, the diffusion coefficients $D_{a1}$ and $D_{h1}$ may have
different values. Figs. 3(a)-3(c) show the steady state patterns for the cases
(i) $D_{a1}\,=0.005\,,\,D_{h1}\,=0.01$, (ii) $D_{a1}\,=0.2\,,\,D_{h1}\,=0.005$ and
(iii) $D_{a1}\,=0.008\,,\,D_{h1}\,=0.2$ . In the first two cases, the stationary pattern
has a wave-like appearance. This Turing pattern is different from the one 
consisting of islands that we have been considering so far. For a (1d) RD system,
the values of diffusion coefficients given in (i) and (ii) correspond to the  
situation when the final steady state is homogeneous \cite{Koch}. This fact is reflected in the
patterns seen in Figs. 3(a) and 3(b), the distribution of the  activator concentration
is homogeneous in the horizontal direction. In the third case, the diffusion 
coefficients are such that Turing patterns are formed in the steady state. Fig.3(c)
shows the usual Turing pattern consisting of islands. Thus with an appropriate
choice of diffusion coefficients, one may generate different types of Turing patterns.

                            Case IV considers the situation of random diffusion
coefficients. The diffusion coefficient $D_{ij}$ for diffusion between a pair of sites
is chosen from a binary distribution. The diffusion coefficients for the activator
and the inhibitor  are 0.005 and 0.2 respectively with a probablity p. The diffusion
coefficients have equal values, 0.005, with probablity (1-p). The diffusion term
is now discretized as

\begin{eqnarray*}
D_a\,\bigtriangleup a\,=\,\sum_j\,\,D_{{\bf a_{jk}}}\,[\,a(j,\,t)\,-\,a(k,\,t)\,] \quad\quad\quad       (5)
\end{eqnarray*}
where k denotes the lattice site $x_{ij}$ and j denotes the four n.n. sites.
When all the $D_{{\bf a_{jk}}}$'s are equal to $D_a$, the original discretization is
recovered. Figs. 4(a)-4(b) show the steady state patterns for $p\,= 0.5,\, 0.3\,  and\, 0.2$ 
respectively. In Fig. 4(a), the Turing activator-concentration peaks are above
the steady state value of 1. Figs. 4(b) shows that Turing patterns are still 
formed but the peaks are above or below the value 2. This implies that there
is an overall activation at all the lattice sites. This is an interesting 
feature of the model considered. Fig. 4(c) shows that at $p\,=0.2$, the steady state
is homogeneous but has a higher concentration (a,h) = (2,2) than in the original
steady state for which (a,h) = (1,1). The transition from a steady state  with 
Turing patterns to a new homogeneous steady state is analogous to a dynamical
phase transition and occurs at a value of p in between $p\,=0.2$ and $p\,=0.3$.

\section*{IV. Discussion}

In Case I, we have studied pattern formation in a square network with missing
connections. In chemical RD systems which exhibit Turing patterns, the RD process
takes place in a gel which consists of an irregularly-connected network of pores
of various diameters through which the molecules diffuse. RD processes in the brain
also occur in an irregular network. The present study shows that disorder in
the underlying network has no adverse effect on the formation of Turing patterns.
This is because the length scales involved in the RD process are small.

In our model we have considered random n.n. connectivity. In a more general 
context, when further-neighbour connections are also present, an interesting 
problem to study is the effect of network connectivity on the stability of
the dynamical system. When small perturbations are applied to a steady state, 
the stability of this state may be studied by linear stability analysis, i.e., 
by Taylor-expanding in the neighbourhood of the steady state. Only the first 
two terms in the expansion are kept, the second term contains the derivative
or Jacobian matrix. The original steady state is stable only if all the eigenvalues
of the Jacobian matrix have negative real parts. Consider a randomly-assembled
dynamical network. Full connectance implies that an element of the network 
is connected to all the other elements. For random connectivity, the Jacobian
matrix has random elements. For such a Jacobian matrix of zero mean, The
Wigner-May Theorem \cite{Wigner,May} states that the dynamical system is almost 
surely unstable if the connectivity exceeds a threshold.
Raghavachari and Glazier \cite{Raghavachari} have considered 1d coupled map
lattices with a scaling form of connectivity. Each pair of sites i and j are
connected with the probablity
\begin{eqnarray*}
p_{ij}\,&=&\,\frac{1}{|\vec{r}_i\,-\,\vec{r}_j|^\alpha}\,\, ,\quad  j \,= \pm 1,\,2,\,3,\,....\quad\quad\quad(6)
\end{eqnarray*}
Where $\vec{r}_i$ and $\vec{r}_j$ are the coordinates of the ith and jth sites,
respectively. The n.n. coupling limit corresponds to  $\alpha\rightarrow\infty$
and $\alpha\rightarrow 0$ is the global coupling limit. For this model, the
Jacobian matrix has all non-negative elements and is unstable for low values of 
connectivity but is stable when connectivity exceeds a critical value. In the light
of these studies, it is of interest to include further-neighbour connections
in the model studied in Case I and  study the effect of the richer connectivity
on the formation of Turing patterns. For this purpose, a discretization scheme
involving an extended neighbourhood can be used. Further-neighbour connectivity may be important
in neural networks in which RD prosses are responsible for self-organization
in the neural activity \cite{Liang}.

In Case II, the model studied involves random dynamics. A possible realization
of this situation is as follows. Autocatalysis of the activator may require
the presence of a chemical molecule or some triggering mechanism not available
at all the lattice sites. The chemical molecule in question can be static with 
diffusion coefficient zero because of a large size. The situation is hypothetical
but not unrealistic.

In Case III and Case IV, we have studied models with anisotropic and random
diffusion coefficients. In ordinary diffusion, molecules move from regions of 
greater to regions of less concentration at a rate proportional to the gradient
of the concentration and also proportional to the diffusivity of the substance.
Normally, the diffusivities are inversely proportional to the square roots of
the molecular weights. However in a porous medium with tortuous geometry, the 
diffusion coefficients may be space dependent. The pores of the cell-walls in
a biological system restrict the movement of large molecules in addition to
that imposed by their weights and most of them are unable to pass through
the walls of the cell. In aqueous solution, the activator and inhibitor molecules
may have the same diffusion coefficients but if the RD system is embedded in a gel as in the experiments \cite{Castets,Lee} to observe Turing patterns, the
activator molecules being larger in size are effectively trapped. This provides 
a big difference in diffusion coefficients of activator and inhibitor. In these
examples, the diffusion coefficient is determined by the geometrical structure of the 
RD medium. In a porous medium with pore sizes distributed over a range, the 
diffusion coefficient may very well be space-dependent.

Recent experimental evidence of Turing patterns in physical, chemical and biological
systems has given rise to renewed interest in the study of these patterns.
In this paper, we have considered the effect of randomness and anisotropy on
the pattern formation process. The studies have been based on computer simulation
and the results obtained show that the Turing mechanism of pattern formation is
fairly robust in the presence of randomness and anisotropy.

\section*{Acknowledgement}
The Authors thank Sitabhra Sinha and Asimkumar Ghosh for computational help.
One of the Authors (IC) is supported by the Council of Scientific and Industrial
Research, India under sanction No9/15(173)/96-EMR-I.

\newpage
\section*{Figure Captions}
\begin{description}
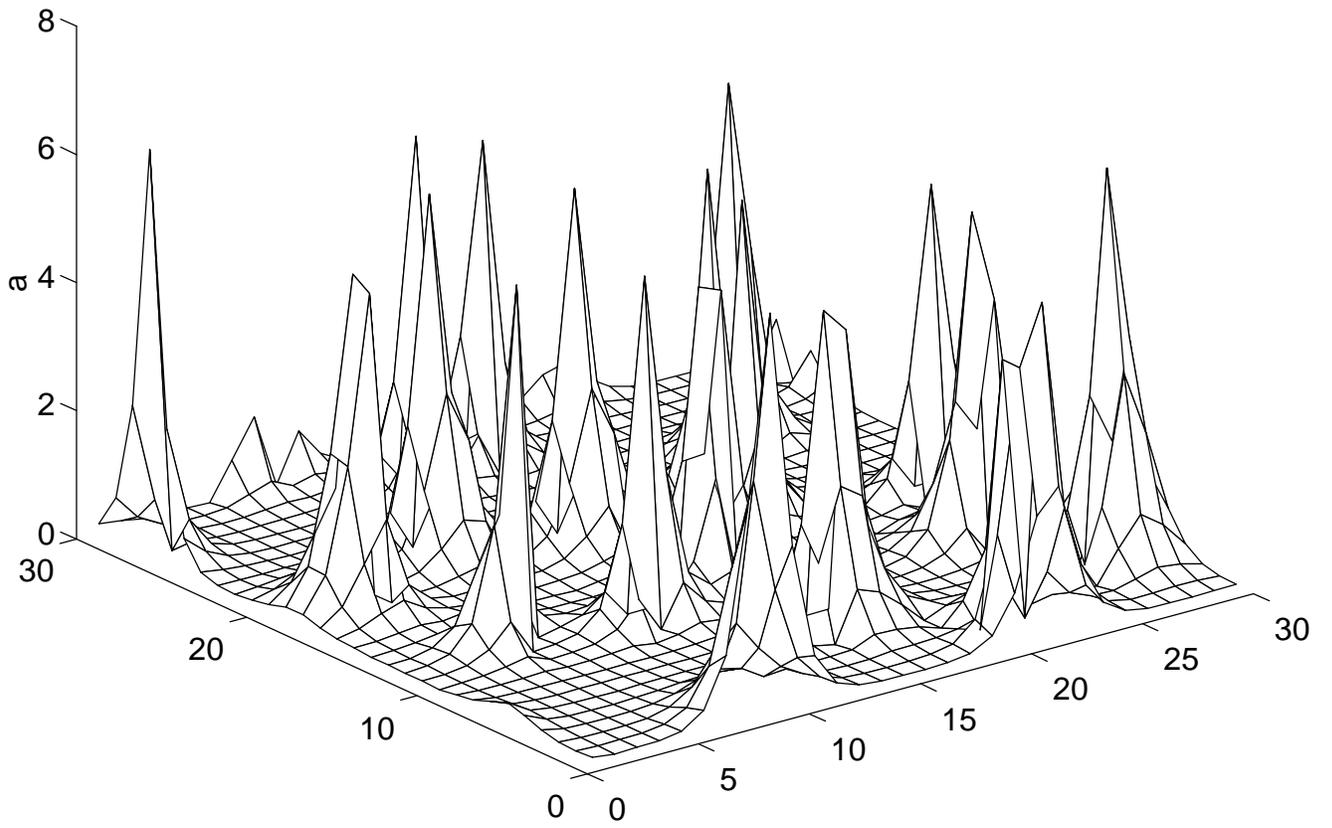
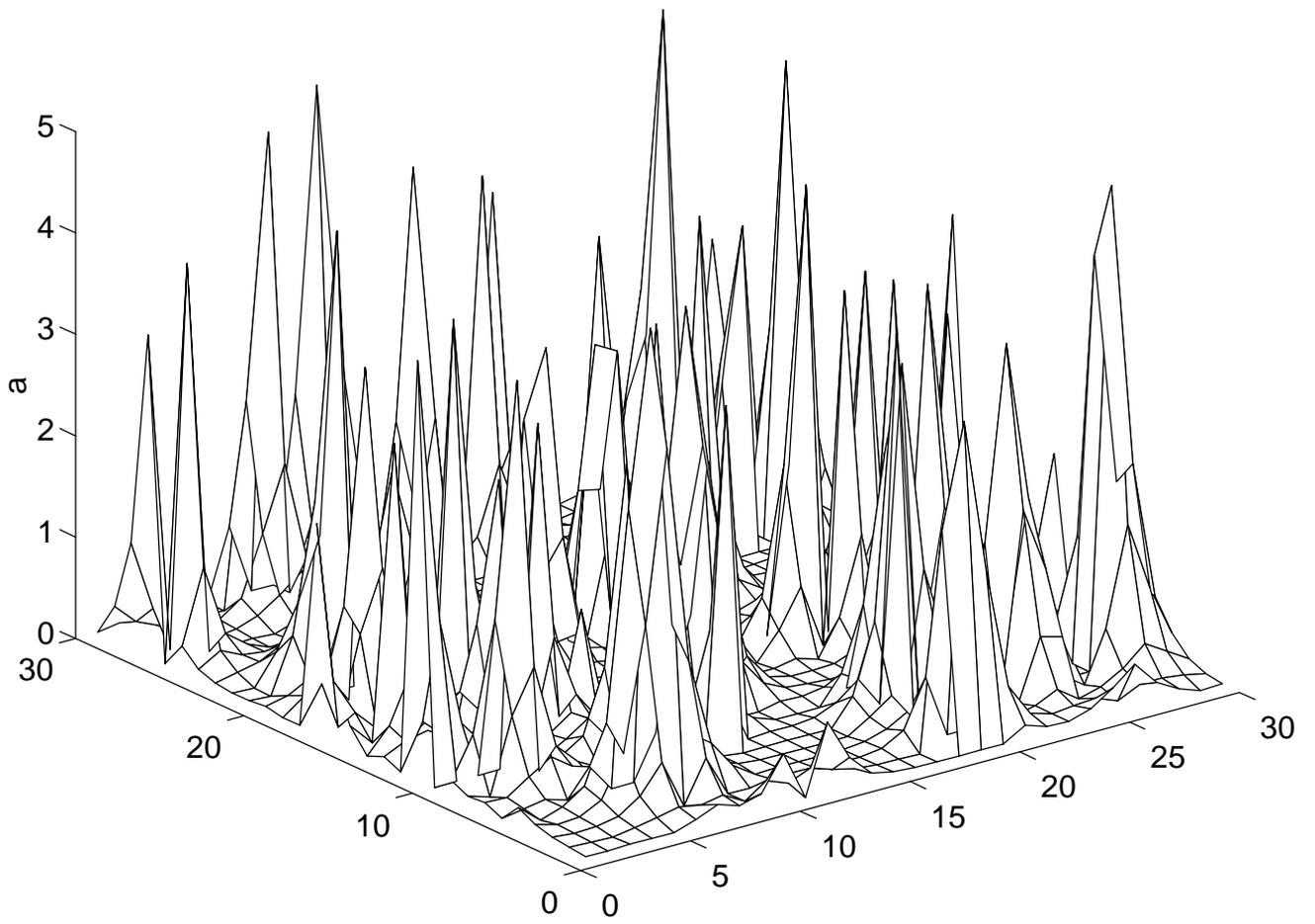
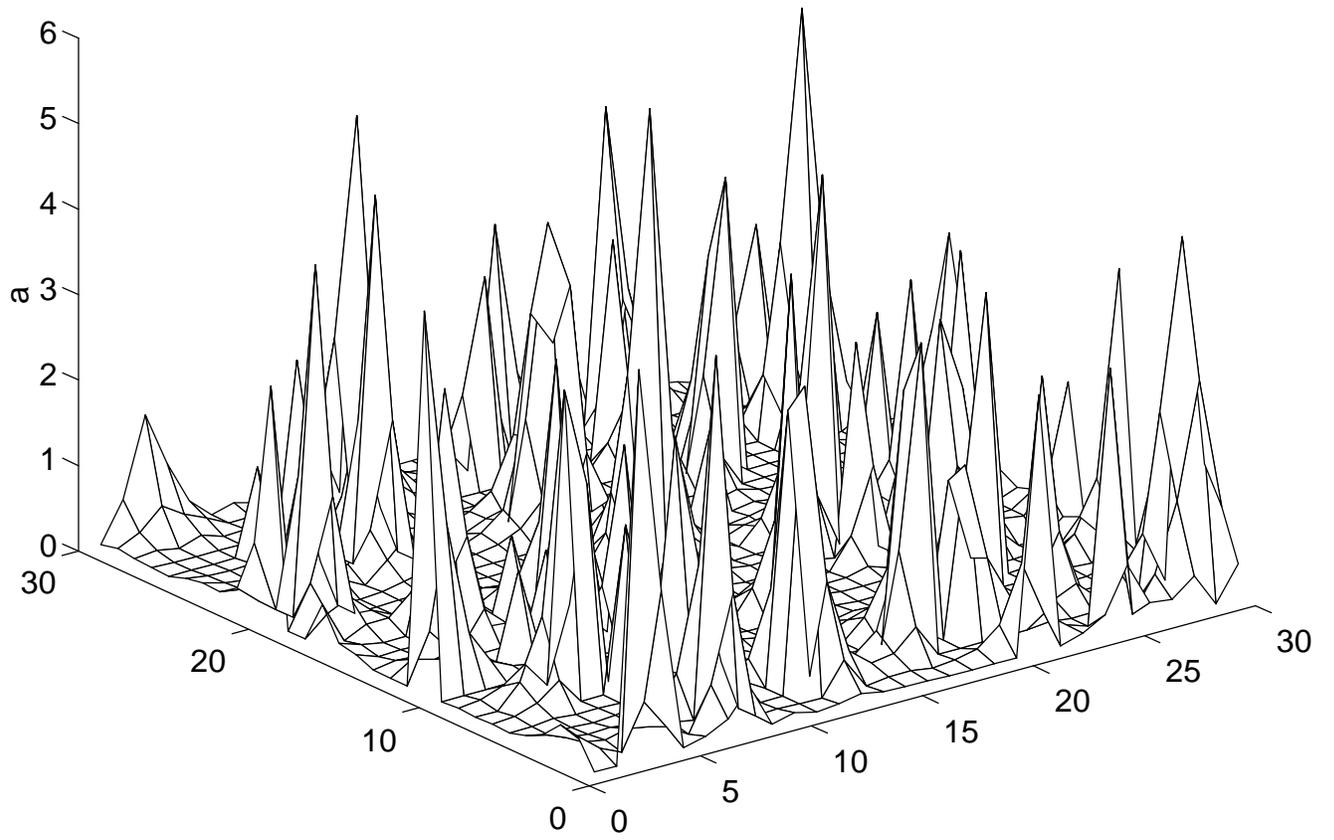
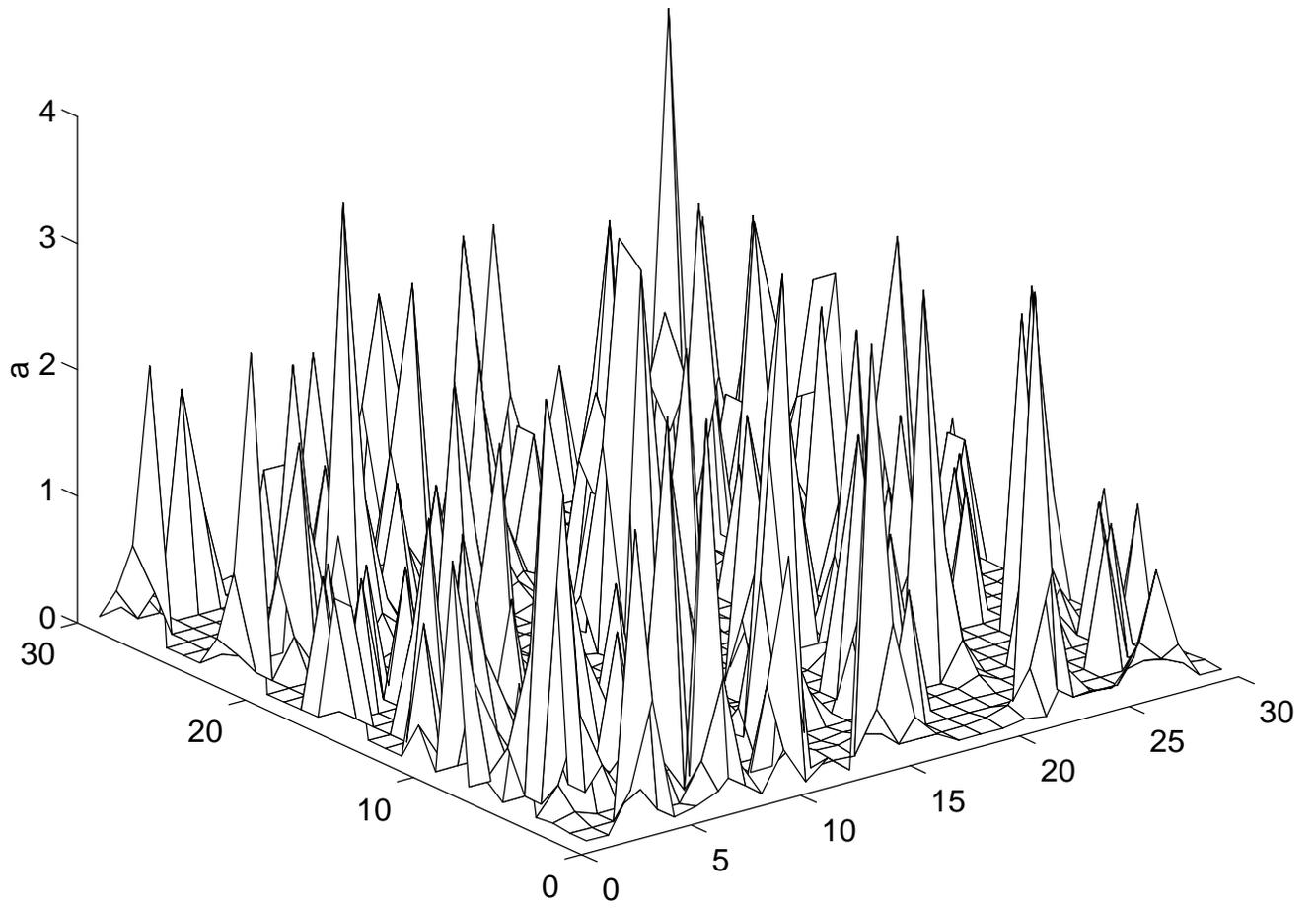
\item[Fig.1] Concentration profiles of the activator on a disordered lattice
structure for site occuption probablities (a) p = 0.9, (b) p = 0.7, (c) p = 0.59
and (d) p = 0.4  (Case I). The islands of high activator concentration constitute
the Turing pattern.
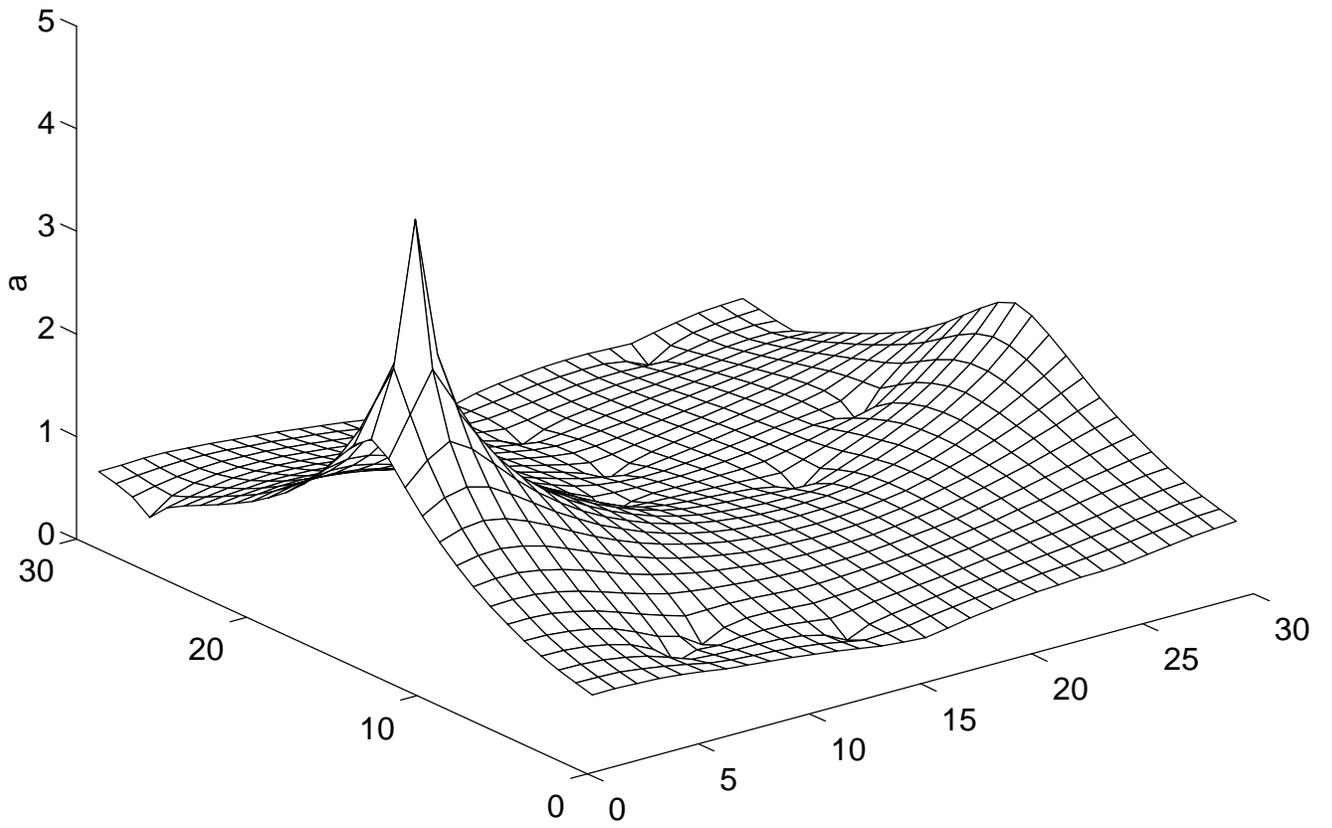
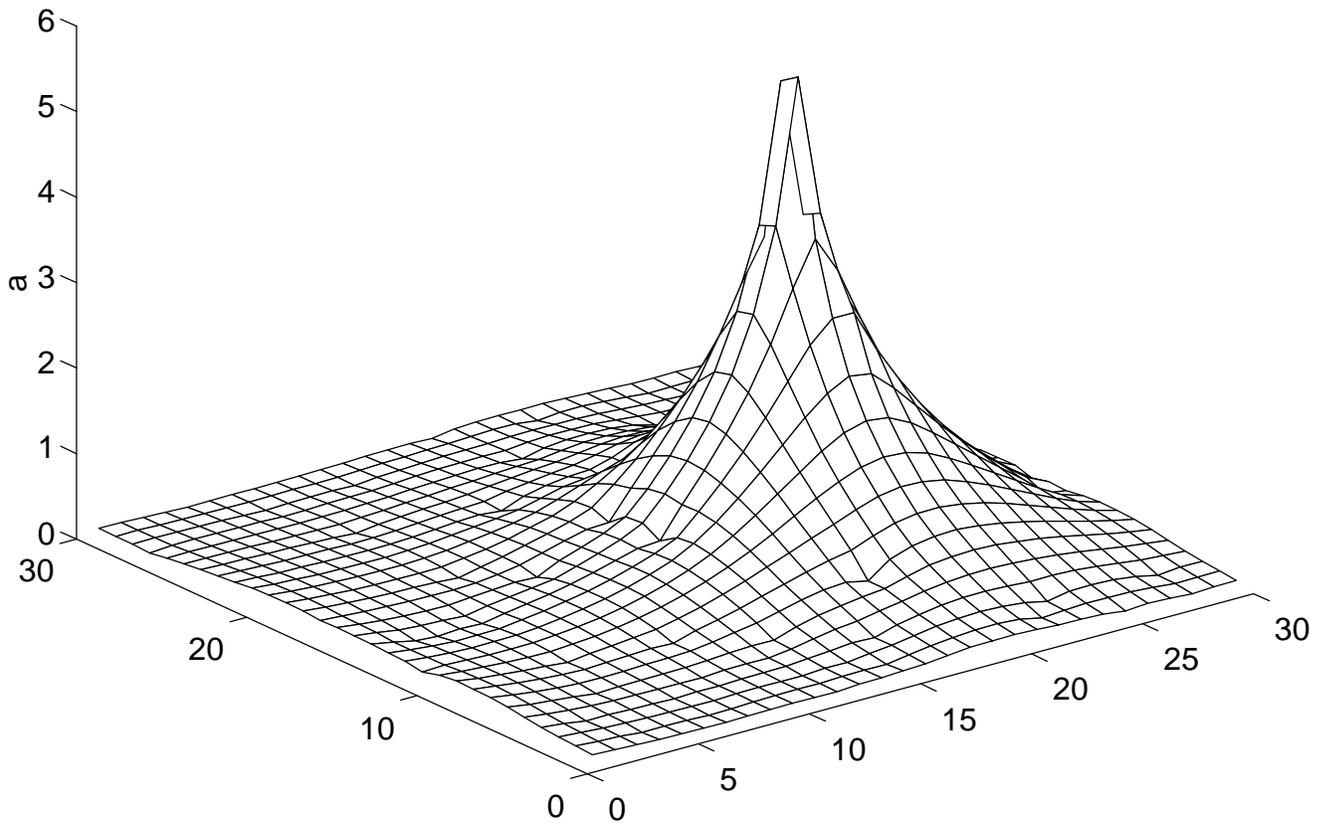
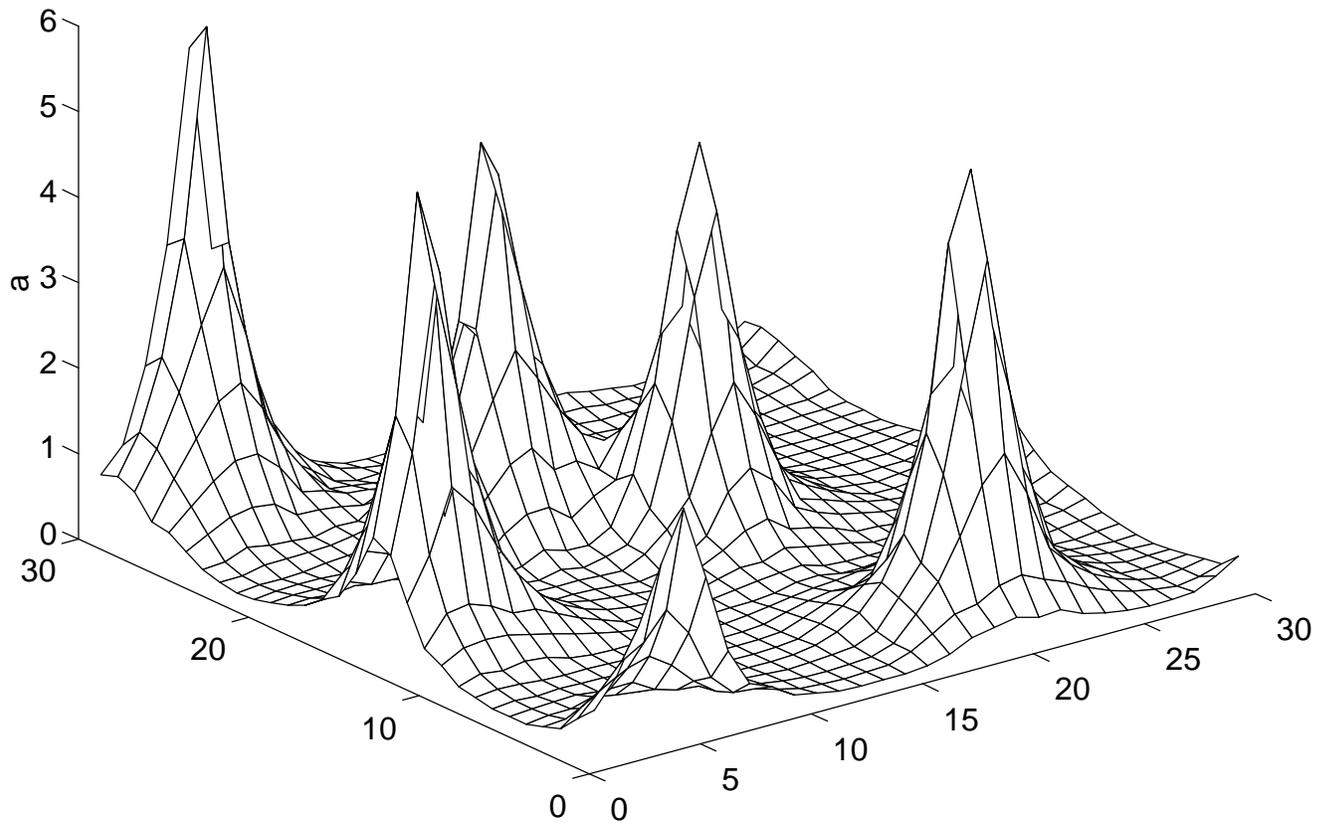
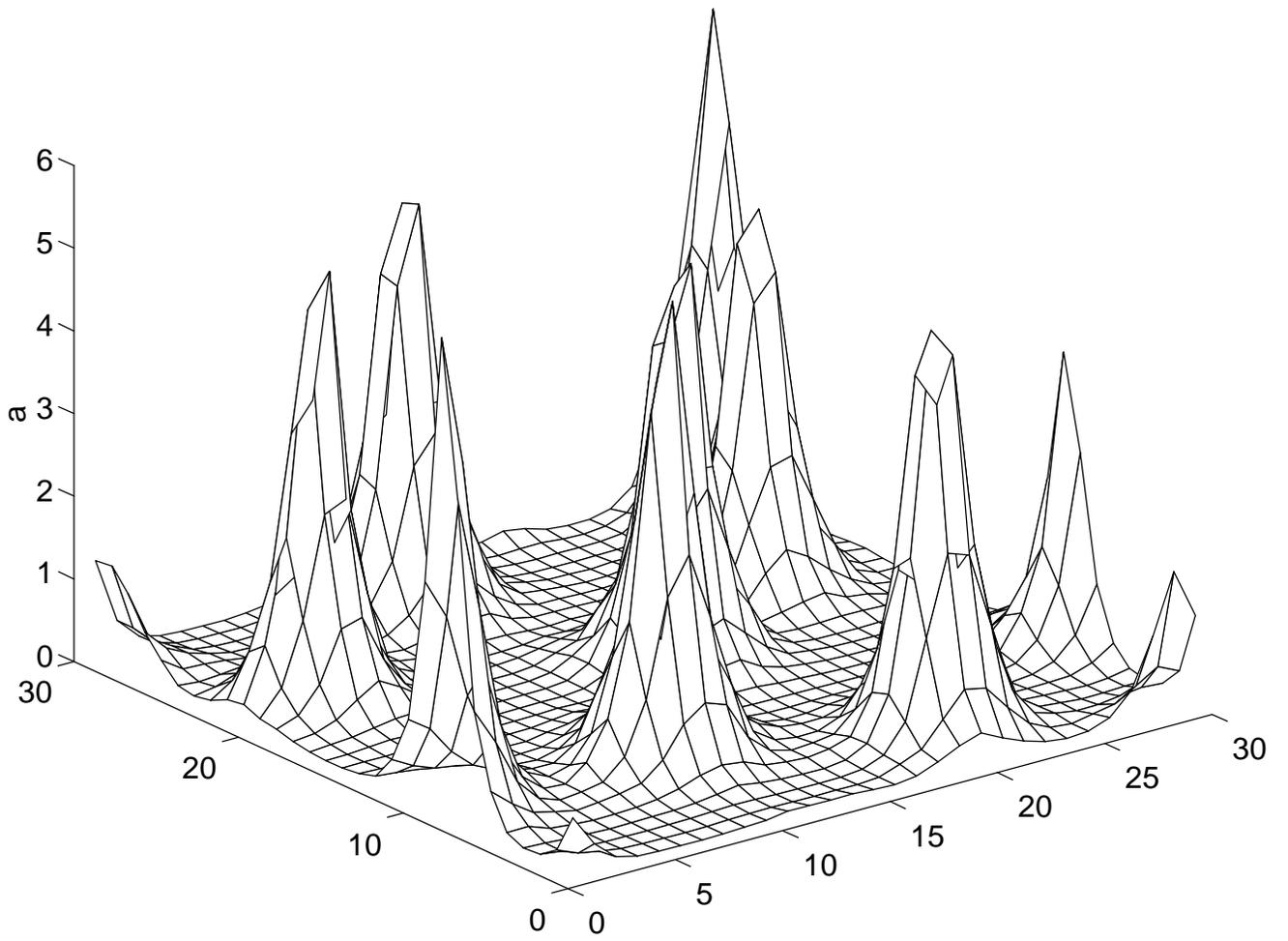
\item[Fig.2] Turing patterns(`a' denotes the activator concentration) for
(a) p = 0.01, (b) p = 0.05, (c) p =0.3 and (d) p = 0.7 (Case II), where p is
the probablity that at a lattice site both reaction and diffusion occur and
(1-p) the probablity that there is only diffusion. The size of the lattice is 
$30\,\times\,30$ .
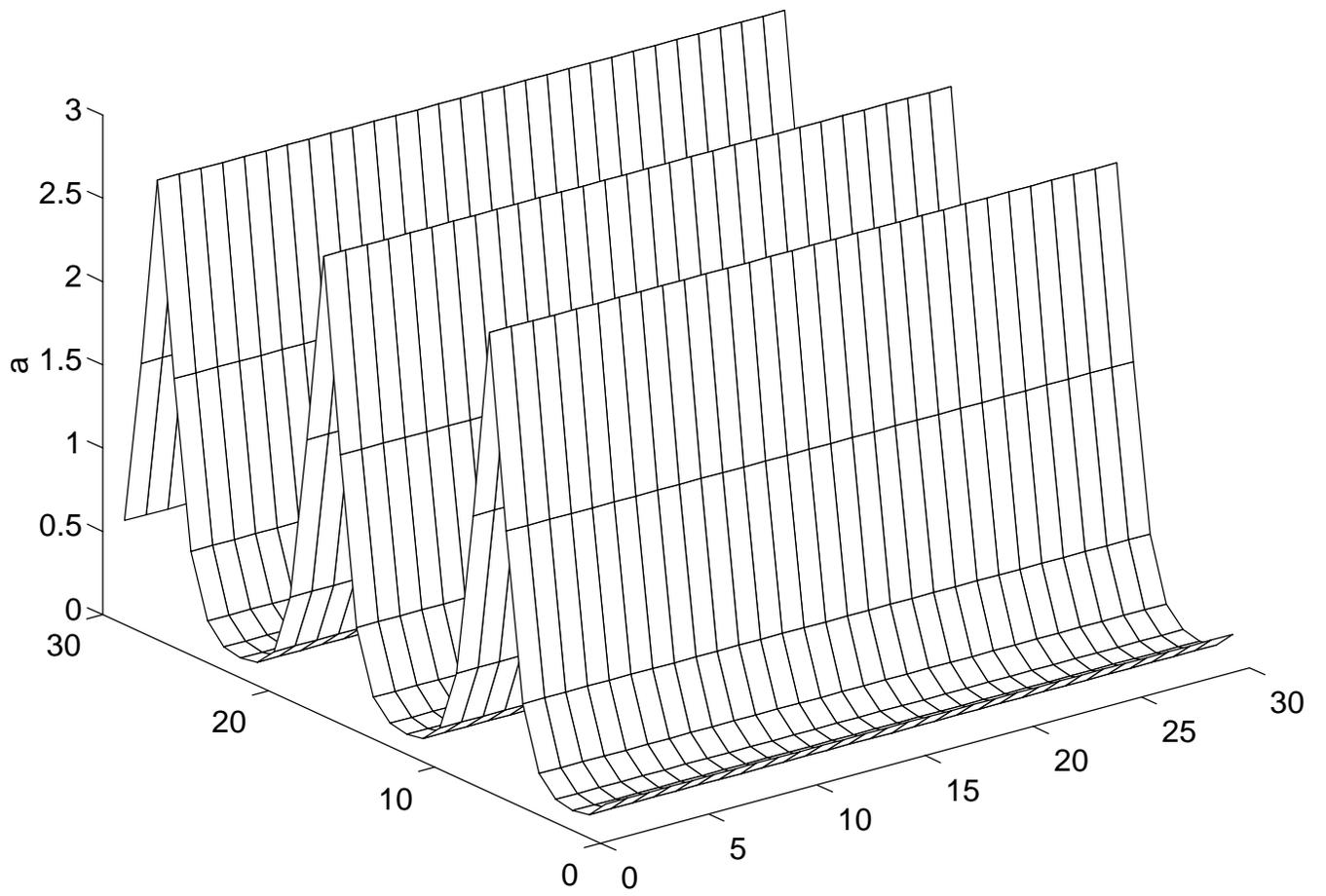
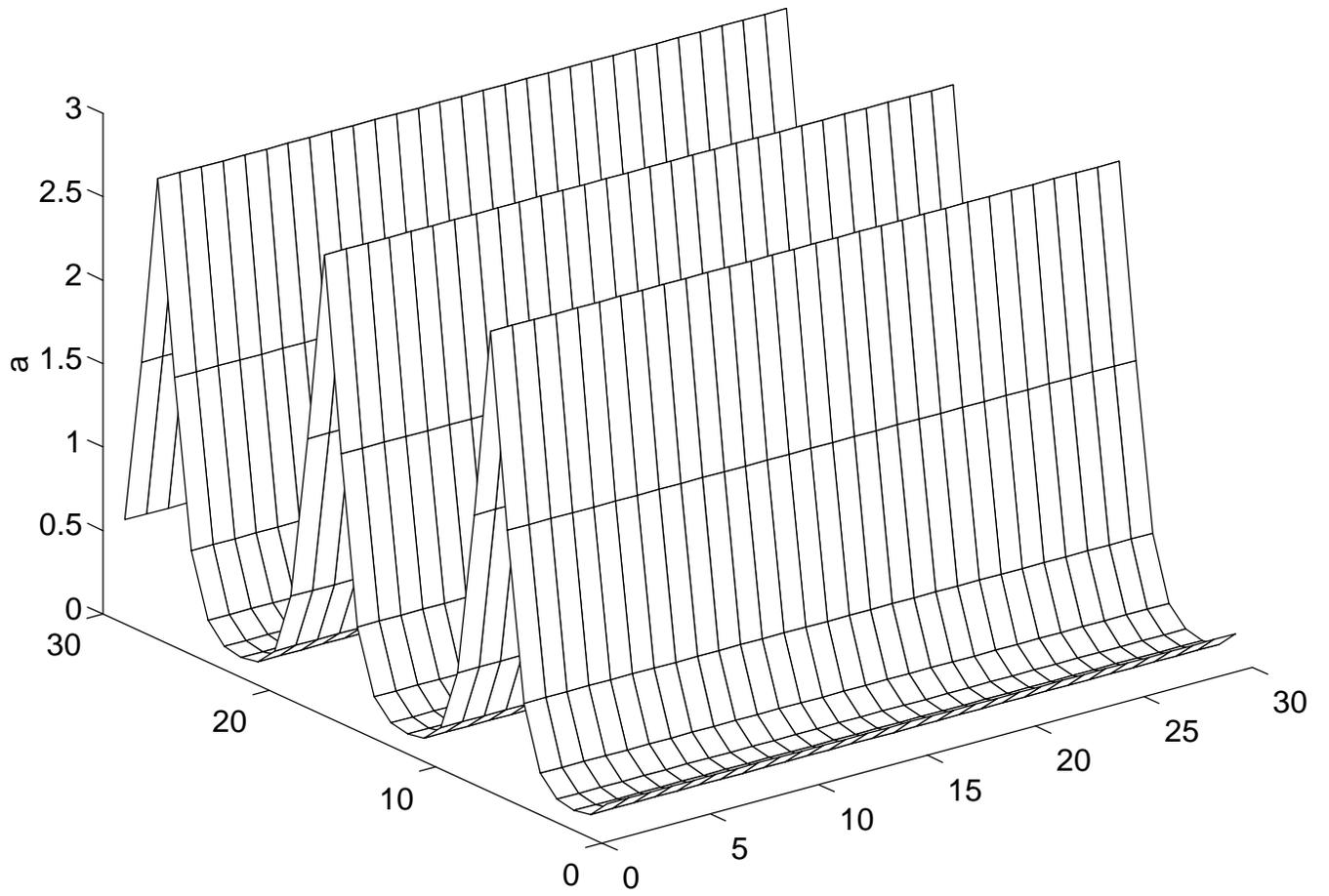
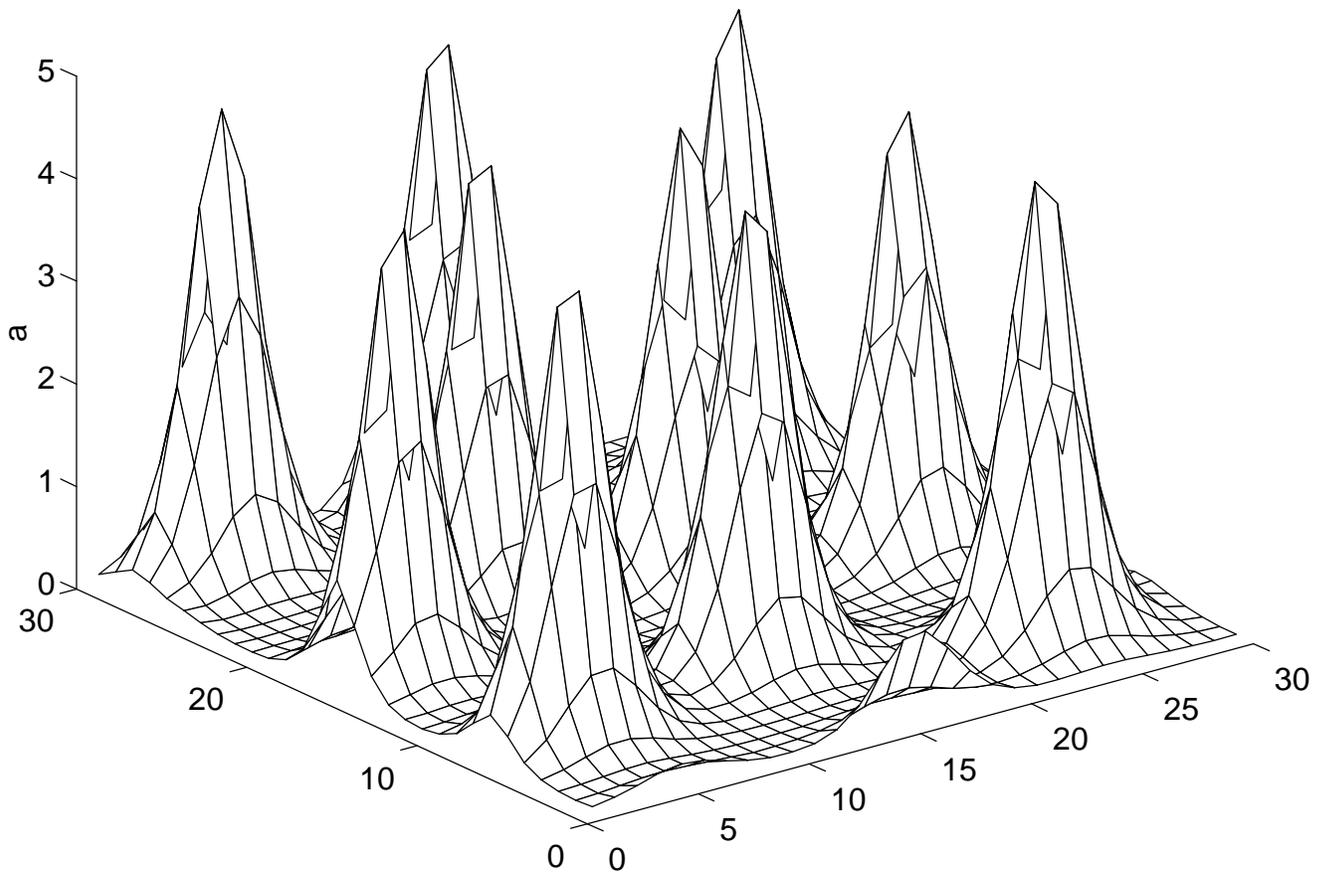
\item[Fig.3] Steady-state patterns in the case of anisotropic diffusion coefficients (Case III). For
diffusion in the vertical direction, the diffusion coefficients of the activator
and inhibitor are: $D_a\,=\,0.005\, and \,D_h\,=\,0.2$. For diffusion in the  
horizontal direction, the diffusion coefficients of the activator and inhibitor
are: (a) $D_{a1}\,=\,0.005,\,D_{h1}\,=\,0.01$ (b) $D_{a1}\,=\,0.2,\,D_{h1}\,=\,0.005$ and (c)
 $D_{a1}\,=\,0.008,\,D_{h1}\,=\,0.2$ .
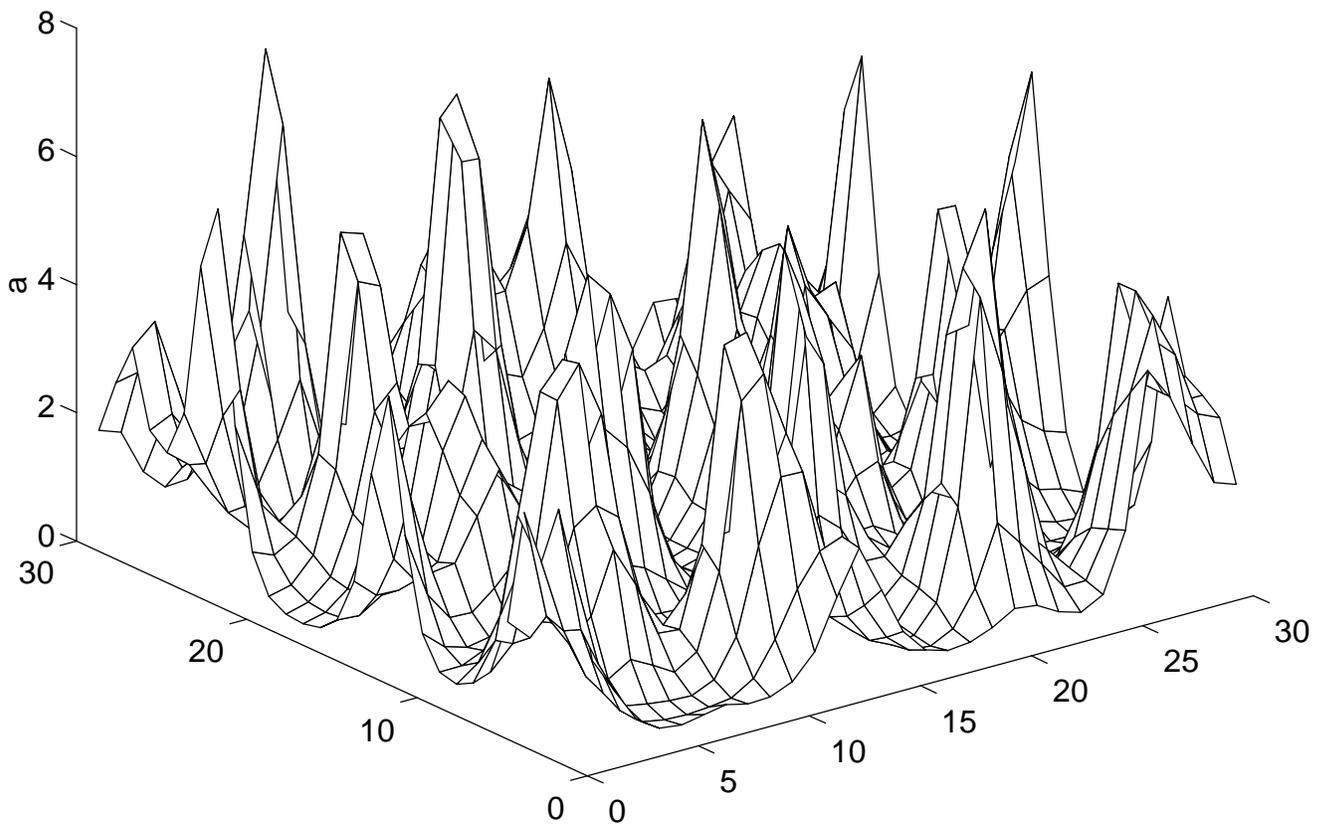
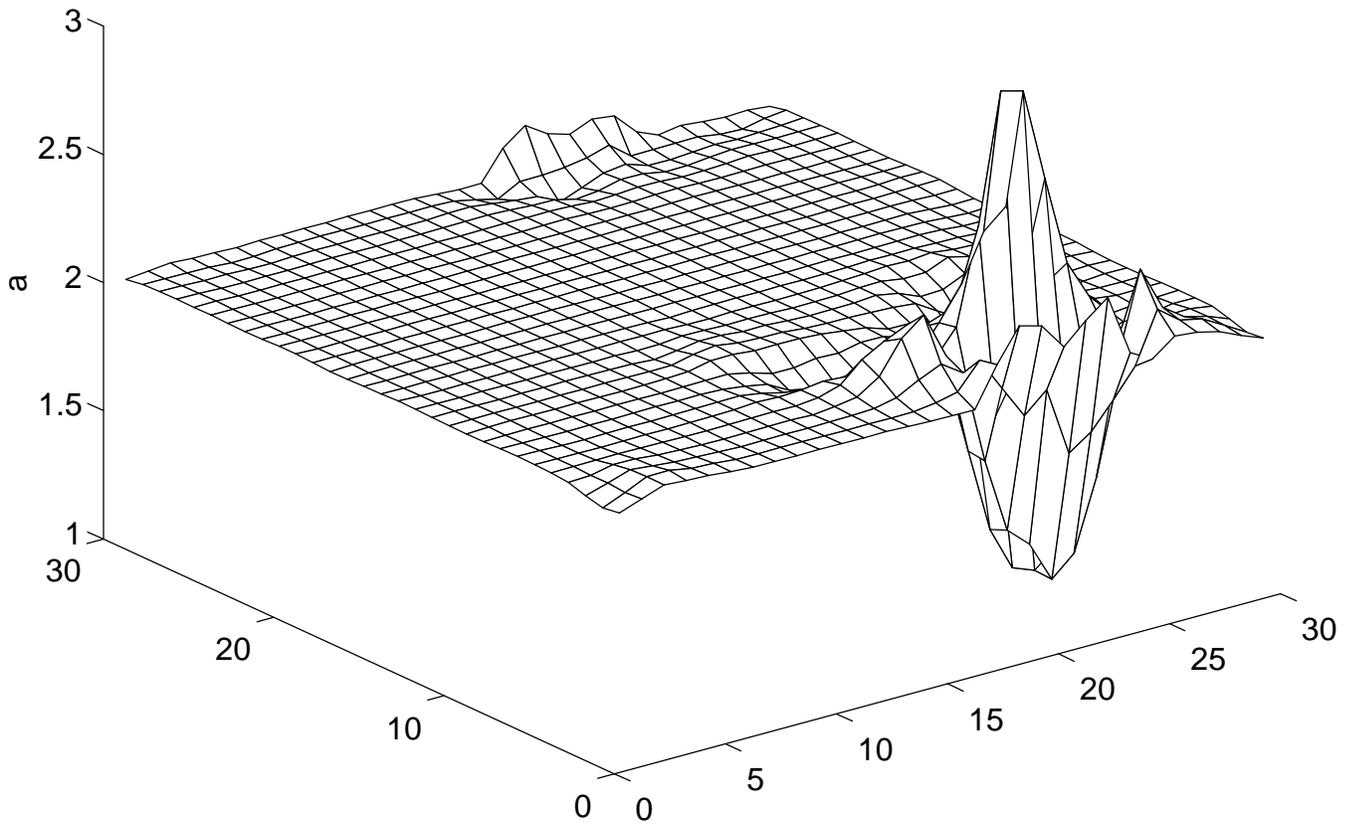
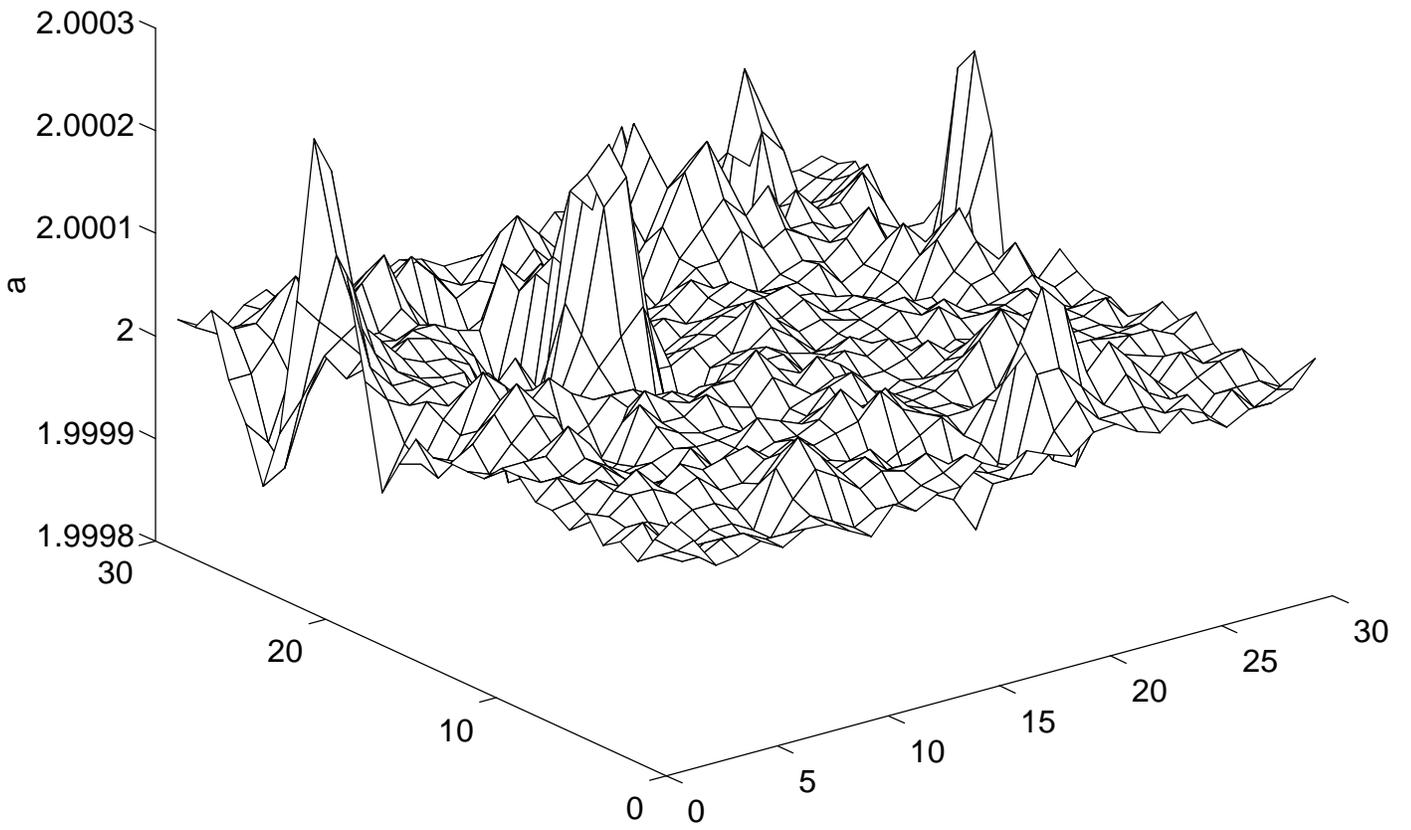
\item[Fig.4] Steady-state pattern for random diffusion coefficients (Case IV).
The diffusion coefficients of the activator and inhibitor are 0.005 and 0.2, respectively, with
a probablity p. The diffusion coefficients have equal values, 0.005, with probablity
(1-p). The values of p that have been considered are: (a) p = 0.5, (b) p = 0.3
and (c) p = 0.2 .
\end{description}

\end{document}